\documentstyle[12pt,epsfig]{article}
\renewcommand{\bar}[1]{\overline{#1}}

\begin{document}
\bigskip\bigskip
\begin{center}
{\large \bf Interference Effects, Time Reversal Violation \\
and Search for New Physics \\
in Hadronic Weak Decays}
\end{center}
\vspace{8pt}
\begin{center}
\begin{large}
  
Z.~J.~Ajaltouni$^{1}$\footnote{ziad@clermont.in2p3.fr},
E.~Di Salvo$^{1,2}$\footnote{Elvio.Disalvo@ge.infn.it}
\end{large} 

\bigskip

$^1$ 
Laboratoire de Physique Corpusculaire de Clermont-Ferrand, \\
IN2P3/CNRS Universit\'e Blaise Pascal, 
F-63177 Aubi\`ere Cedex, France\\

\noindent  
$^2$ 
I.N.F.N. - Sez. Genova,\\
Via Dodecaneso, 33, 16146 Genova, Italy \\  

\noindent  

\vskip 1 true cm
\end{center}
\vspace{1.0cm}

\vspace{6pt}
\begin{center}{\large \bf Abstract}

We propose some methods for studying hadronic sequential two-body 
decays involving more spinning particles. It relies on the analysis 
of T-odd and T-even asymmetries, which are related to interference 
terms. The latter asymmetries turn out to be as useful as the former 
ones in inferring time reversal violating observables; these in turn 
may be sensitive, under some particular conditions, to possible 
contributions beyond the standard model. Our main result is that one 
can extract such observables even after integrating the differential 
decay width over almost all of the available angles. Moreover we find 
that the correlations based exclusively on momenta are quite general, 
since they provide as much information as those involving one or more 
spins. We generalize some methods already proposed in the literature 
for particular decay channels, but we also pick out a new kind of time 
reversal violating observables. Our analysis could be applied, for 
example, to data of LHCb experiment.

\end{center}

\vspace{10pt}

\centerline{PACS numbers: PACS Nos.: 11.30.Er, 11.80.Et, 13.25.-k, 13.30.Eg}

\newpage

\section{Introduction}
Interference is a typical quantum mechanical effect and can 
be exploited experimentally to detect important phenomena. 
For example, parity violation was discovered in the fifties 
by hypothesizing\cite{ly}, and successively exhibiting\cite{wu}, 
an interference term between two decay amplitudes which 
behave differently under parity inversion; these were identified 
later\cite{fg} as matrix elements of the vector and axial current 
respectively. As is well-known, such a term is proportional 
to a scalar product of the type ${\bf J}\cdot{\bf p}$\cite{ly}, 
where ${\bf J}$ and ${\bf p}$ are respectively an angular 
momentum and a momentum of particles involved in the process
under study. It is worth recalling that this dependence was 
picked out by a suitable asymmetry\cite{wu}. Similarly, the 
asymmetries connected to direct CP violation $-$ revealed, it
is worth mentioning, only many years after the discovery of 
indirect CP violation in the $K_L \to \pi\pi$ decays\cite{crf1,crf2} 
$-$ can be detected if two different amplitudes, with different 
weak and strong phases, contribute to the decay\cite{btp,wf1}.  

It could be tempting to propose for time reversal violation 
(TRV) something similar to parity violation. That is, we 
could study interference terms between amplitudes which 
behave differently under time reversal (TR). 
These terms correspond to correlations of the type\cite{va}   
\begin{equation}
C = {\bf v}_1 \cdot {\bf v}_2 \times {\bf v}_3, 
\label{todd}
\end{equation}
where ${\bf v}_i$ ($i$ = 1,2,3) are either momenta or 
angular momenta of particles involved in the process.
Such correlations, named T-odd\cite{wf1,at,der,bi,si}, 
can be defined in processes where more than two particles 
are involved. Correlation (\ref{todd}) may be revealed by 
the asymmetry\cite{va}
\begin{equation}
{\cal A} = \frac{(C>0) - (C<0)}{(C>0) + (C<0)}. 
\label{asm}
\end{equation}
However in this case we are not so lucky as with parity 
violation. Indeed, the genuine TRV effects are
mimicked by fictitious T-odd interference terms, caused 
for example by strong or electromagnetic spin-orbit 
interactions\cite{br1,br2}. Such effects, present in 
scattering processes\cite{der,bi,si,br1,br2} as well as 
in decays (see for example refs. 
\cite{wf1,va,at,d0a,ba1,be2,moo,rlb,els,dt,lnk,bek}), 
produce the same T-odd correlations (\ref{todd}) as real 
TRV. In particular, they mask real TRV in a weak decay, 
owing to final state interactions (FSI). It is therefore 
quite a serious problem to separate the two 
contributions\cite{wf1,va,bdl2}.
But combining the asymmetry of a given triple product 
with the one of its CP-conjugated gives rise to a real 
TRV\cite{va,bdl2,dl}. 

CP violations (CPV), as well as TRV, which can be seen 
as the counterpart of CPV owing to the CPT symmetry, are 
often used for investigating possible clues of new 
physics (NP). Indeed, baryon-antibaryon asymmetry in the
Universe cannot be explained by the sole CKM mechanism of
CPV\cite{gho}. A common strategy to find hints beyond the 
standard model (SM) consists of investigating experimentally
those processes where the CPV (or TRV) are tiny in the SM 
predictions. Indeed some of such violations may indicate 
deviations from the SM. Especially, we recall $B\to\pi K$
decays\cite{bpk1,bpk2,bpk3}, $B_s-{\bar B}_s$ 
mixing\cite{cdf,d0} and the like-sign dimuon asymmetry\cite{abz}. 
However till now experimental and theoretical 
uncertainties do not allow any conclusive results. 
Confirmations to the first few discrepancies are 
demanded. In this sense, experiments on sequential 
two-body decays, mainly of the type 
\begin{equation}
B_{(s)}\to V_1V_2 \label{bvv}
\end{equation}
(with $V_{1,2}$ vector mesons), have been 
performed\cite{be2,bpk2,cdf,d0} and 
suggested\cite{dt,hsv,kli,ch1,ffm,sz,dil,bry}. Such types of 
decays, where more spinning particles are involved, 
 offer the advantage of the angular 
 correlations\cite{wf1,va,bdl2,dl,bdl1,bl1,ddr,dqs}, which may
 provide the moduli and the relative phases of the helicity
 or transversity amplitudes\cite{cw,chn}.
It is worth noting, in the context of such decays, that the above 
mentioned T-odd  correlations are more and more frequently 
proposed\cite{moo,rlb,els,dt,lnk} and also  used\cite{d0a,ba1,be2} for 
discovering possible clues of NP.  This is because these correlations may
be preferred to direct CP-asymmetries in the cases when the 
relative strong phases are negligibly small\cite{dl,cg}. Moreover,
as regards the helicity amplitudes, they are a further, suitable tool 
for testing the SM\cite{dil,bry,ka}, possibly by examining 
fake\cite{dil,ddu}  T-odd observables.  
 
The aim of the present paper is to illustrate some methods for
analyzing two-body hadronic weak decays. In particular we study 
the properties of asymmetries connected to T-odd products 
relative to a sequential two-body decay; moreover, we consider 
some T-even asymmetries, seldom used in the literature but equally 
useful. First of all, we define suitable T-odd and T-even 
correlations, of the type (\ref{todd}) or slightly more complicated, 
and the corresponding asymmetries. Then we show how to use such
asymmetries for extracting observables sensitive to TRV and, possibly,
to NP. In particular, we illustrate some tests for probing clues 
to physics beyond the SM. This work implements in some way a previous 
one\cite{ajd},where we proposed tests based on the knowledge of moduli 
and relative phases of the amplitudes of decays of the type 
considered here. Moreover it generalizes some of the methods 
adopted in the literature in decays of the type (\ref{bvv}).

The paper is organized as follows. Sects. 2 to 4 are devoted to 
the definitions and to the analytical expressions of single and 
double T-odd and T-even asymmetries, written as functions of 
helicity amplitudes. Several TRV observables are defined
in sect. 5. In sect. 6 we carry out an explicit example. 
Sect. 7 is dedicated to relations with previous papers. 
Various tests of the SM are presented in sect. 8. Lastly 
some conclusions are drawn in sect. 9. 

\section{Single T-odd and T-even Asymmetries}

Here and in the following section we illustrate some T-odd 
and T-even asymmetries, which can be inferred from a sequential 
decay of the type 
\begin{equation}
J\to a ~~ b, ~~~~~~ a\to a_1 ~~ a_2, ~~~~~~ b\to b_1 ~~ b_2.
\label{sdc}
\end{equation}
Such observables depend on correlations of the type (\ref{todd})
or similar. The T-odd asymmetries have been proposed by several 
authors, in different contexts. The most recent contributions regard
hadronic\cite{dt,bek,bdl2,dl,kli,dil,bdl1,bl1,arg,aj,nsl} and 
semi-leptonic\cite{blnr,cg0,cgn,alv,cg1} decays, searches for top 
decays\cite{at,avl,hmy} and for new particles\cite{moo,els,lnk,bchk}.
Also some experiments\cite{d0a,ba1,be2} adopt this technique.
On the contrary, T-even asymmetries are not so
frequently suggested in the literature\cite{at,aj,daj1}. 
In order to introduce these observables, we define preliminarily 
suitable reference frames. 

\subsection{Reference Frames}
First of all, we define a canonical frame, at rest with 
respect to the parent resonance $J$: 
\begin{equation}
{\hat{\bf y}} = \frac{{\bf p}_{in}}{|{\bf p}_{in}|},  ~~~~~~  
{\hat{\bf z}} =\frac{{\bf p}_{in}\times{\bf p}_J}
{|{\bf p}_{in}\times{\bf p}_J|},   ~~~~~~ 
{\hat{\bf x}} = {\hat{\bf y}}\times{\hat{\bf z}},
\end{equation}
where ${\bf p}_{in}$ and ${\bf p}_J$ are, respectively, the 
momenta of the initial beam  and of the resonance $J$ in the 
laboratory frame.

On the other hand, the successive decays of the particles $a$ 
and $b$ are more conveniently described in the helicity frames. 
For example, as regards particle $a$, it is given by the 
following three mutually orthogonal unit vectors:
\begin{equation}
{\hat{\bf e}}_L = \frac{{\bf p}_a}{|{\bf p}_a|},  ~~~~~~  
{\hat{\bf e}}_T = \frac{{\hat{\bf z}}\times{\hat{\bf e}}_L}
{|{\hat{\bf z}}\times{\hat{\bf e}}_L|},   ~~~~~~ 
{\hat{\bf e}}_N = {\hat{\bf e}}_T\times{\hat{\bf e}}_L,
\label{unvc}
\end{equation}
${\bf p}_a$ being the momentum of $a$ in the canonical 
frame.
 Analogous definitions hold true for particle $b$.

\subsection{Definition of Single Asymmetries}

\subsubsection{Single T-odd Asymmetries}
 
Consider the scalar product 
\begin{equation}
T_a^N = {\bf p}_{a_1}\cdot{\hat{\bf e}}_N, \label{trpd}
\end{equation}
where ${\bf p}_{a_1}$ is the momentum of the particle $a_1$ in 
the rest frame of $a$. This is a T-odd quantity. Correspondingly, 
we define the T-odd asymmetry
\begin{equation}
A_a^N = \frac{N(T_a^N>0)-N(T_a^N<0)}{N(T_a^N >0)+N(T_a^N <0)},\label{tas}
\end{equation}
where $N(T_a^N > 0)$ $[N(T_a^N <0)]$ is the number of decays 
such that, for a given ${\bf p}_{a_1}$,  the scalar product above is positive 
(negative). Another, independent, T-odd product can be defined as 
\begin{equation}
T_b^N = {\bf p}_{b_1}\cdot{\hat{\bf e}}_N, \label{scp}
\end{equation}
where ${\bf p}_{b_1}$ is the momentum of particle $b_1$ in 
the rest frame of $b$. Obviously, this product induces an 
asymmetry analogous to (\ref{tas}). 

If different than zero, such asymmetries do not imply TRV, 
since, as told, they may derive contributions also from strong 
or/and electromagnetic interactions\cite{wf1,va,br1,br2}. 
However, as we shall see in sect. 5, it is possible to 
define quantities sensitive to such a 
violation\cite{va,bdl2,dl}.

\subsubsection{Single T-even Asymmetries}

The scalar products 
\begin{equation}
T_a^{T(L)} = {\bf p}_{a_1}\cdot{\hat{\bf e}}_{T(L)} \label{tevn}
\end{equation}
are T-even quantities. Analogously to the T-odd 
asymmetries of subsect. 2.2.1, we define the T-even asymmetries
\begin{equation}
A_a^{T(L)} = \frac{N(T_a^{T(L)}>0)-N(T_a^{T(L)}<0)}{N(T_a^{T(L)} >0)+N(T_a^{T(L)} <0)}.\label{evas}
\end{equation}
Also in this case it is possible to define two more (T-even) asymmetries 
by substituting ${\bf p}_{b_1}$ to ${\bf p}_{a_1}$ in the definitions 
(\ref{tevn}).

\subsection{Differential Two-Body Decay Width}

We deduce, here and in the following subsections, the expressions 
of the differential width of a two-body decay and of the asymmetries
just defined before. The calculations are based on the formalism of 
the density matrix\cite{jwk,agp1,mnt,bfr}.  
The starting point is the differential decay width 
for sequential two-body decays \cite{ag,agp2}. It reads as
\begin{eqnarray}
&~& \frac{1}{\Gamma_{ab}\Gamma_{a_1a_2}\Gamma_{b_1b_2}} 
\frac{d^9\Gamma_{a_1a_2b_1b_2}}{d^2\Omega 
d^2\Omega_a d^2\Omega_b dp^2_J dp^2_a dp^2_b} \nonumber
\\
&=& {\cal N}_{Jab}W\sum_{\mu_a,\mu_b}a^{s_a}_{\mu_a} a^{s_b}_{\mu_b}
{\cal T}_{\mu_a\mu_b} (\Omega,\Omega_a,\Omega_b). \label{dif0}
\end{eqnarray}
Here the $\Gamma$'s are the partial decay widths of the decays
(\ref{sdc}); moreover
\begin{eqnarray}
{\cal N}_{Jab} &=& \frac{2s_a+1}{4\pi}\frac{2s_b+1}{4\pi}\frac{2J+1}{4\pi},
\\
 W~~ &=& |B(p^2_J)|^2|B(p^2_a)|^2|B(p^2_b)|^2,  ~~~ \ ~~~~ \ ~~~~
 \label{bw}
\end{eqnarray}
\begin{eqnarray}
{\cal T}_{\mu_a\mu_b}(\Omega,\Omega_a,\Omega_b) = 
\sum_{\lambda_a,\lambda_b}\sum_{\lambda'_a,\lambda'_b}\alpha^J_{\lambda_a\lambda_b}
\alpha^{J*}_{\lambda'_a\lambda'_b}{\cal D}^{s_a*}_{\lambda_a\mu_a}(\Omega_a)
\nonumber
\\
\times {\cal D}^{s_a}_{\lambda'_a\mu_a} (\Omega_a) 
{\cal D}^{s_b*}_{\lambda_b\mu_b}(\Omega_b)
{\cal D}^{s_b}_{\lambda'_b\mu_b}(\Omega_b)
\sum_{M,M'}{\cal D}^{J*}_{M\Lambda}(\Omega)
{\cal D}^J_{M'\Lambda'}(\Omega)\rho^{(0)}_{MM'},
\end{eqnarray}
\begin{eqnarray}
\alpha^J_{\lambda_a\lambda_b} &=& A^J_{\lambda_a\lambda_b}
\large(\sum_{\lambda_a,\lambda_b}
|A^J_{\lambda_a\lambda_b}|^2\large)^{-1/2},
\label{nmam}
\\
a^{s_a}_{\mu_a} &=& \sum_{\mu_{a_1}-\mu_{a_2}=\mu_a}
|A^{s_a}_{\mu_{a_1}\mu_{a_2}}|^2\large(\sum_{\mu_{a_1},\mu_{a_2}}
|A^{s_a}_{\mu_{a_1}\mu_{a_2}}|^2\large)^{-1}, \label{sdch}
\\
a^{s_b}_{\mu_b} &=& \sum_{\mu_{b_1}-\mu_{b_2}=\mu_b}
|A^{s_b}_{\mu_{b_1}\mu_{b_2}}|^2\large(\sum_{\mu_{b_1},\mu_{b_2}}
|A^{s_b}_{\mu_{b_1}\mu_{b_2}}|^2\large)^{-1}.
\end{eqnarray}
$J$, $s_a$ and $s_b$ are the spins, respectively, of the parent
resonance and of particles $a$ and $b$. $A^J_{\lambda_a\lambda_b}$,
$A^{s_a}_{\mu_{a_1}\mu_{a_2}}$ and $A^{s_b}_{\mu_{b_1}\mu_{b_2}}$
are the rotationally invariant two-body decay amplitudes relative
to the three decays (\ref{sdc}). They depend, respectively, on the 
helicities $\lambda_a(\lambda'_a)$ and $\lambda_b(\lambda'_b)$,
$\mu_{a_1}$ and $\mu_{a_2}$, and $\mu_{b_1}$ and $\mu_{b_2}$,
such that 
\begin{equation}
\lambda_a-\lambda_b = \Lambda, ~~~~ \lambda'_a-\lambda'_b =
\Lambda', ~~~~ \mu_{a_1}-\mu_{a_2}=\mu_a, ~~~~ 
\mu_{b_1}-\mu_{b_2}=\mu_b.
\end{equation}
$M (M')$ is the third component of the spin of $J$ in the 
canonical frame. $\Omega$, $\Omega_a$, $\Omega_b$ denote the 
directions of, respectively,  particles $a$, $a_1$ and $b_1$ in 
the  rest frames of their parent resonances: as usual, we have 
set $\Omega \equiv (\theta,\phi)$, where $\theta$ and $\phi$ 
are, respectively, the polar and azimuthal angle; similar
definitions hold for $\Omega_a$ and $\Omega_b$.
$\rho^{(0)}_{MM'}$ and the $B(p^2)$'s are, respectively,
the spin density matrix of the parent resonance and the 
relativistic Breit-Wigner functions of the resonances, 
normalized as
\begin{equation}
\sum_M\rho^{(0)}_{MM} (p^2, p^2) = 1, ~~~~~~ 
\int_0^{\infty}dp^2|B(p^2)|^2 = 1. \label{nrml}
\end{equation}
Lastly, ${\cal D}^J_{M\Lambda}(\Omega)$ is a Wigner
${\cal D}$-matrix element, defined as\cite{edms} 
\begin{eqnarray}
{\cal D}^J_{M\Lambda}(\Omega) &=& e^{-iM\phi}d^J_{M\Lambda}
(\theta),
\\
d^J_{M\Lambda}(\theta) &=& \langle JM|e^{-iJ_y\theta}
|J\Lambda\rangle, \label{dfc}
\end{eqnarray}
where $J_y$ is the $y$-component of the angular momentum.

Problems arise if one takes into account the energy dependence 
of the decay amplitudes and of the density matrix; however, if the 
resonances are sufficiently narrow, as it usually happens, such a 
dependence can be neglected. Therefore, in an analysis, it would 
be convenient to integrate over the energies of the resonances. But 
we do not perform such  integrations, in order to point out 
an important theoretical subtlety, to be clarified in sect. 5. 

Now we integrate eq. (\ref{dif0}) over $d^2\Omega_a d^2\Omega_b$. 
To this end, we have to take into account the normalization
\begin{equation} 
\int d^2\Omega {\cal D}^{J*}_{M\Lambda}(\Omega)
{\cal D}^J_{M'\Lambda}(\Omega) = \frac{4\pi}{2J+1}\delta_{MM'}
\label{norm}  
\end{equation}
for Wigner's ${\cal D}$-matrices. As a result we get 
the expression of the differential two-body decay width,  {\it i. e.},
\begin{equation}
\Gamma(\Omega)=
{\cal N}_{J}W\sum_{\lambda_a,\lambda_b} |\alpha^J_{\lambda_a\lambda_b}|^2
\sum_{M,M'}{\cal D}^{J*}_{M\Lambda}(\Omega)
{\cal D}^J_{M'\Lambda}(\Omega)\rho^{(0)}_{MM'}.
\label{didw} 
\end{equation}
Here 
\begin{equation}
{\cal N}_{J} = \frac{2J+1}{4\pi}.
\end{equation}
Rigorously speaking, the distribution $\Gamma$ depends also on $p_J^2$,
$p_a^2$ and $p_b^2$, through the factor $W$. However, in order to simplify 
the notation, we  have omitted such a dependence in the 
argument of the distribution. We shall adopt this convention 
for all of the asymmetries that we shall define in the present 
paper. 

More important, we remark that the differential decay width (\ref{dif0}), and therefore all 
of the expressions derived from it, do not depend strictly on the decay amplitudes, 
but rather on the dimensionless quantities (\ref{nmam});  as we shall see, these play an 
important role in the definitions of the asymmetries  and, from now on, will be named "reduced" amplitudes.

\subsection{Analytical Expression of the T-odd Asymmetry} 

The expression of the observable $\Gamma(\Omega) 
A_a^N(\Omega)$ reads
\begin{eqnarray}
\Gamma(\Omega)A_a^N(\Omega) &=&  \int_0^{\pi}
 d\theta_a sin\theta_a \Big[\int_{-\pi/2}^{\pi/2}-\int_{\pi/2}^{3\pi/2}\Big] d\phi_a \int 
d^2\Omega_b \Delta\Gamma (\Omega, \Omega_a \Omega_b) \nonumber
\\
&=&{\cal N}_{Ja} W
\sum_{\mu_a}a^{s_a}_{\mu_a} \sum_{\lambda_a,\lambda_b}\sum'_{\lambda'_a}
 \Theta^{s_a}_{\lambda_a \lambda'_a\mu_a}\frac{4(-)^{D_a}}{\lambda_a-\lambda'_a}
\nonumber
\\
&\times&\alpha^J_{\lambda_a\lambda_b}\alpha^{J*}_{\lambda'_a\lambda_b}
\sum_{M,M'}{\cal D}^{J*}_{M\Lambda}(\Omega)
{\cal D}^J_{M'\Lambda'}(\Omega)\rho^{(0)}_{MM'}.\label{ntash}  
\end{eqnarray}
Here $\Delta\Gamma (\Omega, \Omega_a \Omega_b)$ is a 
short notation for the differential decay width 
(\ref{dif0}). Moreover we have set
\begin{eqnarray}
{\cal N}_{Ja} &=& \frac{2J+1}{4\pi} \frac{2s_a+1}{4\pi},  ~~~~   
D_a = (\lambda_a-\lambda'_a-1)/2,
\\
\Theta^{s_a}_{\lambda_a\lambda'_a\mu_a} &=& \int_0^{\pi}d\theta_a 
sin\theta_a d^{s_a}_{\lambda_a\mu_a}(\theta_a) d^{s_a}_{\lambda'_a\mu_a}(\theta_a), ~~~~ \ 
~~~~
\label{thetl} 
\end{eqnarray}
the $d$-matrices being defined by eq. (\ref{dfc}).
Lastly, we have denoted by $\sum'$ the sum over those 
$\lambda'_a$ for which $\lambda_a -\lambda'_a$ is odd.

\subsection{Expressions of the T-even Asymmetries}

The expression of the asymmetry $ A_a^T$ is given by 
 
\begin{eqnarray}
\Gamma(\Omega) A_a^T(\Omega) &=& \int_0^{\pi} sin\theta_a 
d\theta_a \Big[\int_0^{\pi}-\int_{\pi}^{2\pi}\Big] d\phi_a 
\int d^2\Omega_b \Delta\Gamma (\Omega, \Omega_a \Omega_b)
\nonumber\\
&=& {\cal N}_{Ja} W \sum_{\mu_a}
a^{s_a}_{\mu_a} \sum_{\lambda_a,\lambda_b}\sum'_{\lambda'_a}
 \Theta^{s_a}_{\lambda_a \lambda'_a\mu_a}\frac{4i}{\lambda_a-\lambda'_a}
\nonumber
\\
&\times&\alpha^J_{\lambda_a\lambda_b}\alpha^{J*}_{\lambda'_a\lambda_b}
\sum_{M,M'}{\cal D}^{J*}_{M\Lambda}(\Omega)
{\cal D}^J_{M'\Lambda'}(\Omega)\rho^{(0)}_{MM'}.\label{ttash}  
\end{eqnarray}
On the other hand, the asymmetry $ A_a^L$ reads 
\begin{eqnarray}
\Gamma(\Omega) A_a^L(\Omega) &=& 
\Big[\int_0^{\pi/2}-\int_{\pi/2}^{\pi}\Big] sin\theta_a d\theta_a 
\int_0^{2\pi} d\phi_a \int d^2\Omega_b
\Delta\Gamma (\Omega, \Omega_a \Omega_b) \nonumber
\\
&=& 8\pi{\cal N}_{Ja} W\sum_{\mu_a>0}
\Delta a^{s_a}_{\mu_a} 
\nonumber
\\
&\times& \sum_{\Lambda}  \Delta a^J_{\Lambda\mu_a}
\sum_{M,M'}{\cal D}^{J*}_{M\Lambda}(\Omega)
{\cal D}^J_{M'\Lambda}(\Omega)\rho^{(0)}_{MM'}.\label{ltash}  
\end{eqnarray}
Here
\begin{eqnarray}
\Delta a^{s_a}_{\mu_a} &=& \frac{1}{2}(a^{s_a}_{\mu_a}- a^{s_a}_{-\mu_a}),
\label{dlts}\\
\Delta a^J_{\Lambda\mu_a} &=& \frac{1}{2}\sum_{\lambda_a > 0}
\delta^{s_a}_{\lambda_a\mu_a}(|\alpha^J_{\lambda_a\lambda_b}|^2-
|\alpha^J_{-\lambda_a\lambda'_b}|^2),
\label{ffd}
\end{eqnarray}
$\lambda_b = \lambda_a-\Lambda$, $\lambda'_b = -\lambda_a-\Lambda$ and 
\begin{equation}
\delta^{s_a}_{\lambda_a\mu_a} = 
\Big[\int_0^{\pi/2}-\int_{\pi/2}^{\pi}\Big]
[d^J_{\lambda_a\mu_a}(\theta)]^2sin\theta d\theta.
\end{equation}

\subsection{Remarks}
Two short remarks are in order. Firstly, the expressions of 
the single asymmetries
calculated in this section for particle $a$ can be extended in a
straightforward way to particle $b$. Secondly, we observe that 
the procedure just described, concerning single asymmetries,
is not applicable to a decay of the type (\ref{bvv}), but only to cases
where both the parent resonance 
and at least one of the decay products are spinning 
($J, s_a (s_b)\geq 1/2$); for example, it may be applied to 
\begin{equation}
\Lambda_b \to \Lambda(\Lambda_c) ~~ P(V), ~~~~~ \ ~~~~~
\Lambda_c \to \Lambda ~~ÊP(V), \label{declb}
\end{equation}
$P$ and $V$ denoting, respectively, a pseudoscalar and a vector 
particle. 

\section{Double T-odd and T-even Asymmetries}

Here we define some double asymmetries, analogous to the single
asymmetries of the previous section. As an example, we set
\begin{equation}
A_{ab}^{NT} = \frac{\Big[N(T_a^N\cdot T_b^T>0)-N(T_a^N\cdot T_b^T<0)\Big]}
{\Big[N(T_a^N\cdot T_b^T>0)+N(T_a^N\cdot T_b^T<0)\Big]}. \label{NTas}
\end{equation}
It is instructive to notice that the numerator of this asymmetry can be written as
\begin{eqnarray} 
{\mathrm Num} &=& \Big[N(T_a^N>0, T_b^T>0)-N(T_a^N>0, T_b^T<0)\Big]\nonumber
\\
~~~~ &-& \Big[N(T_a^N<0, T_b^T>0)-N(T_a^N<0, T_b^T<0)\Big],
\end{eqnarray}
which justifies the name of ''double'' asymmetry.
This is a T-odd asymmetry, as appears from eq. 
(\ref{NTas}). Another asymmetry of this type can be defined as   
\begin{equation}
A_{ab}^{LN} = \frac{\Big[N(T_a^L\cdot T_b^N>0)-N(T_a^L\cdot T_b^N<0)\Big]}
{\Big[N(T_a^L\cdot T_b^N>0)+N(T_a^L\cdot T_b^N<0)\Big]}.\label{NLas} 
\end{equation}
Two more T-odd asymmetries can be obtained, respectively, 
from eqs. (\ref{NTas}) and (\ref{NLas}) by interchanging 
$a$ with $b$. 

{\it Viceversa}, one can define five T-even asymmetries, 
in a quite analogous way: $A_{ab}^{NN}$, $A_{ab}^{TT}$, 
$A_{ab}^{LL}$, $A_{ab}^{TL}$ and $A_{ab}^{LT}$. For
example,
\begin{equation}
A_{ab}^{NN} = \frac{\Big[N(T_a^N\cdot T_b^N>0)-N(T_a^N\cdot T_b^N<0)\Big]}
{\Big[N(T_a^N\cdot T_b^N>0)+N(T_a^N\cdot T_b^N<0)\Big]}.\label{NNas} 
\end{equation}

Analytically, the first double, T-odd asymmetry reads as 
\begin{eqnarray}
&~& \Gamma(\Omega)A_{ab}^{NT}(\Omega) \ ~~~~~~~~ \ ~~~~~~~~ \ ~~~~~~~~ \ ~~~~~~~~ \
\nonumber
\\
&=& \Big[\int_{-\pi/2}^{\pi/2}-\int_{\pi/2}^{3\pi/2}\Big] d\phi_a
\Big[\int_0^{\pi}-\int_{\pi}^{2\pi}\Big] d\phi_b 
\int_0^{\pi}sin\theta_a d\theta_a
\int_0^{\pi}sin\theta_b d \theta_b \Delta\Gamma(\Omega, \Omega_a \Omega_b) \nonumber
\\
&=& {\cal N}_{Jab} W \sum_{\mu_a,\mu_b}
a^{s_a}_{\mu_a} a^{s_b}_{\mu_b}\sum_{\lambda_a,\lambda_b} \sum'_{\lambda'_a}\sum'_{\lambda'_b} 
\Theta^{s_a}_{\lambda_a\lambda'_a\mu_a} 
\Theta^{s_b}_{\lambda_b\lambda'_b\mu_b}
\frac{16i(-)^{D_a}}{(\lambda_a-\lambda'_a) (\lambda_b-\lambda'_b)}
\nonumber
\\
&\times&\alpha^J_{\lambda_a\lambda_b}\alpha^{J*}_{\lambda'_a\lambda'_b}
\sum_{M,M'}{\cal D}^{J*}_{M\Lambda}(\Omega)
{\cal D}^J_{M'\Lambda'}(\Omega)\rho^{(0)}_{MM'}.
\label{NTan}
\end{eqnarray}

The other double, T-odd asymmetry is given by
\begin{eqnarray}
\Gamma(\Omega)A_{ab}^{LN}(\Omega) &=& 2\pi{\cal N}_{Jab} W \sum_{\mu_a,\mu_b}
a^{s_a}_{\mu_a} a^{s_b}_{\mu_b}\sum_{\lambda_a,\lambda_b}\sum'_{\lambda'_b} 
\delta^{s_a}_{\lambda_a \mu_a}\Theta^{s_b}_{\lambda_b\lambda'_b\mu_b}
\frac{4(-)^{D_b}}{\lambda_b-\lambda'_b}\nonumber
\\
&\times&\alpha^J_{\lambda_a\lambda_b}\alpha^{J*}_{\lambda_a\lambda'_b}
\sum_{M,M'}{\cal D}^{J*}_{M\Lambda}(\Omega)
{\cal D}^J_{M'\Lambda'}(\Omega)\rho^{(0)}_{MM'},\label{LNan}
\end{eqnarray}

with $D_b = (\lambda_b-\lambda'_b-1)/2$. Concerning the double, T-even asymmetries, we have

\begin{eqnarray}
&~&\Gamma(\Omega)A_{ab}^{NN}(\Omega) \ ~~~~~~~~ \ ~~~~~~~~ \ ~~~~~~~~ \ \nonumber
\\ 
&=& {\cal N}_{Jab} W  \sum_{\mu_a,\mu_b}
a^{s_a}_{\mu_a} a^{s_b}_{\mu_b}\sum_{\lambda_a,\lambda_b} \sum'_{\lambda'_a} 
\sum'_{\lambda'_b} \Theta^{s_a}_{\lambda_a\lambda'_a\mu_a} 
\Theta^{s_b}_{\lambda_b\lambda'_b\mu_b}
\frac{16(-)^{D_a}(-)^{D_b}}{(\lambda_a-\lambda'_a)(\lambda_b-\lambda'_b)}
\nonumber
\\
&\times&\alpha^J_{\lambda_a\lambda_b}\alpha^{J*}_{\lambda'_a\lambda'_b}
\sum_{M,M'}{\cal D}^{J*}_{M\Lambda}(\Omega)
{\cal D}^J_{M'\Lambda'}(\Omega)\rho^{(0)}_{MM'}; \ ~~~~~~~~ \
\label{NNan}
\\
&~&\Gamma(\Omega)A_{ab}^{TT}(\Omega) \ ~~~~~~~~ \ ~~~~~~~~ \ ~~~~~~~~ \ ~~~~~~~~ \
\nonumber
\\ 
&=& {\cal N}_{Jab} W  \sum_{\mu_a,\mu_b}
a^{s_a}_{\mu_a} a^{s_b}_{\mu_b}\sum_{\lambda_a,\lambda_b}\sum'_{\lambda'_a} 
\sum'_{\lambda'_b}\Theta^{s_a}_{\lambda_a\lambda'_a\mu_a} \Theta^{s_b}_{\lambda_b\lambda'_b\mu_b}
\nonumber
\\
&\times&\frac{(-16)}{(\lambda_a-\lambda'_a)(\lambda_b-\lambda'_b)}
\alpha^J_{\lambda_a\lambda_b}\alpha^{J*}_{\lambda'_a\lambda'_b}
\sum_{M,M'}{\cal D}^{J*}_{M\Lambda}(\Omega)
{\cal D}^J_{M'\Lambda'}(\Omega)\rho^{(0)}_{MM'};
\label{TTan}
\end{eqnarray}
\begin{eqnarray}
\Gamma(\Omega)A_{ab}^{LL}(\Omega) &=& 16\pi^2{\cal N}_{Jab} W  \sum_{\mu_a>0} 
\sum_{\mu_b>0}\Delta a^{s_a}_{\mu_a}\Delta a^{s_b}_{\mu_b} \sum_{\Lambda}
\Delta^{(2)}_{\Lambda\mu_a\mu_b}
\nonumber
\\
&\times& \sum_{M,M'}{\cal D}^{J*}_{M\Lambda}(\Omega)
{\cal D}^J_{M'\Lambda}(\Omega)\rho^{(0)}_{MM'}.
\label{LLan}
\end{eqnarray}
Here 
\begin{equation}
\Delta^{(2)}_{\lambda_b \mu_a\mu_b} = \sum_{\lambda_a}
|\alpha^J_{\lambda_a\lambda_b}|^2\delta^{s_a}_{\lambda_a\mu_a}
\delta^{s_b}_{\lambda_a\mu_b},
\end{equation}

with $\lambda_b = \lambda_a-\Lambda$. Lastly,
\begin{eqnarray}
\Gamma(\Omega)A_{ab}^{LT}(\Omega) &=& 2\pi{\cal N}_{Jab} W 
\sum_{\mu_a,\mu_b}a^{s_a}_{\mu_a} a^{s_b}_{\mu_b}
\sum_{\lambda_a,\lambda_b} \delta^{s_a}_{\lambda_a\mu_a} 
\sum'_{\lambda'_b} \Theta^{s_b}_{\lambda_b\lambda'_b\mu_b}
\frac{4i}{\lambda_b-\lambda'_b}
\nonumber
\\
&\times&\alpha^J_{\lambda_a\lambda_b}\alpha^{J*}_{\lambda_a\lambda'_b}
\sum_{M,M'}{\cal D}^{J*}_{M\Lambda}(\Omega)
{\cal D}^J_{M'\Lambda'}(\Omega)\rho^{(0)}_{MM'}.
\label{LTan}
\end{eqnarray}

The expressions of the asymmetries $A_{ab}^{TN}$, 
$A_{ab}^{NL}$ and $A_{ab}^{TL}$ are obtained from, 
respectively, $A_{ab}^{NT}$, $A_{ab}^{LN}$ and 
$A_{ab}^{LT}$ by interchanging particle $a$ with $b$.

The asymmetries defined and elaborated in this 
section may be applied to a wider class of 
decays, including those of the type (\ref{bvv}). 

\section{Summary}

We have shown in the two previous sections that the
expressions of the differential decay widths and of the 
T-odd and T-even asymmetries depend on interference 
terms or on the moduli squared of the "reduced" decay 
amplitudes. Linear combinations of such parameters can be
obtained from experimental data, as we show in
Appendix for several cases of interest. 
Indeed, we apply there the method of the 
moments\cite{ag,agp2} to the distributions defined in sects.
2 and 3. As is well-known, each moment is factorized into a
term which depends solely on the production density matrix,
times another one which contains information only on decay
amplitudes. Moreover, for each moment, the former factor is 
independent of the distribution considered; therefore the ratio
of, say, a given moment of some asymmetry, to  the corresponding
moment of the differential decay width (\ref{didw}), provides
linear relations among the moduli squared or among the 
interference terms of the "reduced" amplitudes. Therefore we 
obtain linear systems with 
respect to these quantities, and constraints to 
be imposed on the parameters which can be 
extracted from data. The results of the analysis 
that we have proposed can be used for determining 
TRV observables, as we shall show in the 
next section. In particular, the linear system 
obtained may be over-determined, provided all 
asymmetries defined in sects. 2 and 3 are nonzero.
Unfortunately, if particle $a$ 
or $b$ (or both) in (\ref{sdc}) have a strong or 
electromagnetic decay, some asymmetries may vanish
because of parity conservation, as we shall see 
in a specific example (sect. 6). However
we shall establish that, in the decay considered, 
the nonzero 
distributions are sufficient to determining all moduli and 
relative phases of the amplitudes. In cases when this is 
possible, these quantities can be used as inputs for the 
tests of NP suggested in our preceding paper\cite{ajd}. 

\section{Determining TRV Observables}

Some TRV observables are deduced directly from experimental distributions,
while others require some elaboration. In this section we shall consider both
kinds of observables, starting from the former ones. 

\subsection{Some TRV Observables from Distributions}

Consider the differential width of the sequential two-body 
decay CP-conjugated to (\ref{sdc}), {\it i. e.},
\begin{equation}
{\bar J} \to {\bar a} ~~ {\bar b}, ~~ \ ~~ {\bar a}\to 
{\bar a_1} ~~ {\bar a_2}, ~~ \ ~~  {\bar b}\to 
{\bar b_1} ~~ {\bar b_2}.\label{cdy}
\end{equation}
 The direction corresponding to $\Omega$ is 
\begin{equation}
{\bar \Omega} \equiv (\pi-\theta, \pi+\phi). \label{sang}
\end{equation}
Consider the T-odd asymmetries for ${\bar J}$. As an example, 
first of all, we elaborate the expression of the single T-odd asymmetry 
${\bar A}_a^N({\bar \Omega})$. It is defined through
\begin{eqnarray}
{\bar\Gamma}({\bar\Omega}){\bar A}_a^N({\bar \Omega}) &=& 
{\cal N}_{Ja} {\bar W} \sum_{\mu_a}
{\bar a}^{s_a}_{-\mu_a} \sum_{\lambda_a,\lambda_b}\sum'_{\lambda'_a} 
\Theta^{s_a}_{-\lambda_a -\lambda'_a-\mu_a} \frac{4(-)^{D_a} }
{\lambda'_a-\lambda_a}
\nonumber
\\
&\times& {\bar \alpha}^J_{-\lambda_a-\lambda_b}
{\bar\alpha}^{J*}_{-\lambda'_a-\lambda_b}
\sum_{M,M'}{\cal D}^{J*}_{M-\Lambda}({\bar \Omega})
{\cal D}^J_{M'-\Lambda'}({\bar \Omega}){\bar \rho}^{(0)}_{MM'}.\label{ntcon}  
\end{eqnarray}
Here ${\bar W}$ is analogous to the factor (\ref{bw}) and 
${\bar \rho}^{(0)}_{MM'}$ is the spin density matrix of the 
resonance ${\bar J}$. If the spin of the parent resonances is 0, 
it is natural to define a TRV observable as\cite{va,dt,dl,kli,dil} 
\begin{equation}
\Delta_a^N = \Gamma(\Omega) A_a^N (\Omega)+
{\bar \Gamma}({\bar \Omega}) {\bar A}_a^N ({\bar \Omega}), \label{trvv}
\end{equation}
where the $+$ sign in eq. (\ref{trvv}) is a consequence of 
the change of sign under CP reflection in the scalar product 
(\ref{trpd}). For a nonzero spin of $J$, $\Delta_a^N$ is a TRV 
observable only  if the production process of 
${\bar J}$ is CP-conjugated to the one of  $J$; 
this implies a relation between the density matrices of the 
two resonances, practically impossible to realize. However, 
as we shall see, we can equally derive useful TRV quantities 
from the experimental data. To this end, we consider an ideal 
experiment, in which we assume that the production process of 
the parent resonance $J$ and the decays of particles $a$ and 
$b$ are invariant under CP reflection. 
Moreover we suppose the production processes of $J$ and 
${\bar J}$ to be described by the same density matrix. Lastly, 
we take account of the CPT theorem, which yields ${\bar W}$ 
= $W,$ and of the following properties of the ${\cal D}$- 
and $d$-matrices:
\begin{equation}
{\cal D}^{J*}_{M-\Lambda}({\bar \Omega}){\cal D}^J_{M'-\Lambda'}({\bar \Omega})
= {\cal D}^{J*}_{M\Lambda}(\Omega){\cal D}^J_{M'\Lambda'}(\Omega),
\end{equation}
\begin{equation}
\Theta^{s_a}_{-\lambda_a -\lambda'_a-\mu_a}  = 
-\Theta^{s_a}_{ \lambda_a\lambda'_a\mu_a} ~~ {\mathrm for} 
~~ \lambda_a-\lambda'_a ~~ {\mathrm odd}.
\end{equation}
As a result, we get
\begin{eqnarray}
{\bar \Gamma}({\bar \Omega}){\bar A}_a^N ({\bar \Omega}) &=& 
-{\cal N}_{Ja} W \sum_{\mu_a}
a^{s_a}_{\mu_a} \sum_{\lambda_a,\lambda_b}\sum'_{\lambda'_a} 
\Theta^{s_a}_{\lambda_a \lambda'_a\mu_a} \frac{4(-)^{D_a}}
{\lambda_a-\lambda'_a}
\nonumber 
\\
&\times& {\bar \alpha}^J_{-\lambda_a-\lambda_b}
{\bar \alpha}^{J*}_{-\lambda'_a-\lambda_b}
\sum_{M,M'}{\cal D}^{J*}_{M\Lambda}(\Omega)
{\cal D}^{J}_{M'\Lambda'}(\Omega)\rho^{(0)}_{MM'}.\label{ntcn}  
\end{eqnarray}
Substituting eqs. (\ref{ntash}) and (\ref{ntcn}) into (\ref{trvv}), we get
\begin{equation}
\Delta_a^N = 4{\cal N}_{Ja} W
\sum_{\lambda_a,\lambda_b}\sum'_{\lambda'_a}\sum_{M,M'}
{\cal C}^{M,M'}_{\lambda_a,\lambda_b,\lambda'_a}
{\cal D}^{J*}_{M\Lambda}(\Omega)
{\cal D}^{J}_{M'\Lambda'}(\Omega). \label{thts}  
\end{equation}
Here
\begin{equation}
{\cal C}^{M,M'}_{\lambda_a,\lambda_b,\lambda'_a} = 
\frac{(-)^{D_a}}{\lambda_a-\lambda'_a}\rho^{(0)}_{MM'} T_{\lambda_a,\lambda'_a}
\cdot\epsilon_{\lambda_a,\lambda_b,\lambda'_a}, \ ~~~~ \ ~~~~ \ 
T_{\lambda_a,\lambda'_a} = \sum_{\mu_a} a^{s_a}_{\mu_a} 
\Theta^{s_a}_{\lambda_a\lambda'_a\mu_a}\label{tht3}  
\end{equation}
and
\begin{equation}
\epsilon_{\lambda_a,\lambda_b,\lambda'_a} = 
\alpha^J_{\lambda_a\lambda_b}\alpha^{J*}_{\lambda'_a\lambda_b}-{\bar 
\alpha}^J_{-\lambda_a-\lambda_b}
{\bar \alpha}^{J*}_{-\lambda'_a-\lambda_b},
\label{epsi}  
\end{equation}
with $\lambda'_a-\lambda_a$ odd.
If $\Delta_a^N$ is different from zero, one has TRV.
But this implies that at least one of the $\epsilon_{\lambda_a,\lambda_b,\lambda'_a}$'s is 
non-vanishing. Although deduced under particular 
assumptions on the production process and on the decay 
of particle $a$, this condition depends solely on the 
decays of $J$ and of its 
anti-particle, which are independent of the production 
process and of the successive decays of $a$ and $b$. Therefore 
the $\epsilon_{\lambda_a,\lambda_b,\lambda'_a}$'s are TRV 
observables for odd $\lambda'_a-\lambda_a$. 

One can define quite similarly the quantities $\Delta_b^N$,
$\Delta_{ab}^{NL}$, $\Delta_{ab}^{TN}$ and those obtained 
by interchanging, respectively, $N$ with $L$ and $T$ with $N$.
Obviously, analogous procedures can be applied to these 
observables: $\Delta_{ab}^{NL}$ gives again rise to the 
difference (\ref{epsi}), while $\Delta_{ab}^{TN}$ yields
\begin{equation}
\epsilon_{\lambda_a,\lambda_b,\lambda'_a\lambda'_b} = 
\alpha^J_{\lambda_a\lambda_b}
\alpha^{J*}_{\lambda'_a\lambda'_b}-{\bar 
\alpha}^J_{-\lambda_a-\lambda_b}
{\bar \alpha}^{J*}_{-\lambda'_a-\lambda'_b},
\label{epsi3}  
\end{equation}
for odd $\lambda'_a-\lambda_a$ and $\lambda'_b-\lambda_b$.

It is worth noting that eqs. (\ref{epsi}) and (\ref{epsi3}) are
both TRV and CP-odd, independent of the CPT symmetry.

\subsection{More TRV Observables}

More TRV observables can be obtained from the moduli and relative 
phases of the decay amplitudes. To this end it is convenient to pass 
from the helicity representation to the $l-s$ one, where $l$ is the 
orbital angular momentum and $s$ the overall spin of the two-particle 
state. In particular, interesting TRV quantities may be defined as 
\begin{equation}
\varepsilon^J_{lss'} =
\Im(\alpha^J_{ls}\alpha^{J*}_{l+1s'}+{\bar\alpha}^J_{ls} {\bar\alpha}^{J*}_{l+1s'}). \label{lsc}
\end{equation}
Here
\begin{equation}
\alpha^J_{ls} = \sum_{\lambda_a\lambda_b}C^{s~~l~J}_{\Lambda ~0 ~\Lambda} ~~
C_{\lambda_a ~-\lambda_b ~\Lambda}^{s_a~~s_b~~s} ~~\alpha^J_{\lambda_a\lambda_b}
\end{equation}
and the $C$'s are the usual Clebsch-Gordan coefficients. The
$\varepsilon^J_{lss'}$, which can be inferred from information on
moduli and relative phases, generalize the TRV observable proposed, {\it e. g.}, 
in refs. \cite{va,dl}.

\subsection{Remarks}

We conclude this section with some remarks, in part connected 
to the CPT symmetry.

First of all, the observables defined in this section are 
expressed as functions of the "reduced" amplitudes (\ref{nmam}). 
This is not the same as obtained, {\it e. g.}, in refs. 
\cite{va,bdl2,dl}, where the very amplitudes are involved.  
Therefore our analysis has picked out new T-odd and TRV 
observables, different from the analogous observables defined 
in those references.
 
Secondly, the CPT symmetry implies that the TRV observables just defined 
are of the type 

\begin{equation}
\sum_k \phi_k R_k,
\end{equation}

where the $R_k$'s are finite, T-even quantities and the $\phi_k$'s 
are phases causing time reversal violation.

Thirdly, it sometimes happens that the weak phase of a decay amplitude is 
negligibly small in comparison with the strong one\cite{ddu}. It is more 
convenient in these cases to consider T-odd quantities 
analogous to eqs. (\ref{epsi}) or (\ref{epsi3}), but with the 
+ sign in place of the - sign, or of the type (\ref{lsc}), but with 
the - sign between the two terms. These observables are invariant 
under TR and, according to the CPT symmetry, also CP-even; they are 
called fake T-odd. We shall see in sect. 8 that they may be 
employed in some tests, as already proposed\cite{dil,ddu}.

Fourthly, as regards the T-even asymmetries, consider 
differences of the type
\begin{equation}
\Delta_a^T = \Gamma(\Omega) A_a^T (\Omega)-
{\bar \Gamma}({\bar \Omega}) {\bar A}_a^T ({\bar \Omega}), 
\label{trvt}
\end{equation}
where the same notations and assumptions as in eqs. 
(\ref{sang}) to (\ref{trvv}) and (\ref{ntcn}) have
been adopted. If significantly different than zero,
these differences may consist, either of possible CPT-odd terms 
and products, or of a fake T-odd amplitude times a real 
TRV one. If CPT symmetry is assumed, these observables 
may be used to set constraints on decay models.

Lastly, it is worth spending some words on studies of CPT violation,
a very hot topic at present\cite{khj1,khj2}. Indeed, the possibility of 
such a violation $-$ connected to Lorentz invariance violation $-$ was 
considered more than ten years ago by Coleman and Glashow\cite{cg01,cg02},
who suggested experiments of neutrino oscillations.  
Among the most recent contributions, we mention the
MiniBooNE\cite{mbn} and Minos\cite{mns1,mns2} experiment, theoretical 
speculations\cite{dlm,gms} and the numerous references cited 
therein. Incidentally, neutrino oscillations offer also the possibility
of studying CPV and TRV in the leptonic sector\cite{khj1,khj2}. Moreover, intrinsic CPT violations, if any, have to 
be disentangled from fake 
violations, induced by neutrino-matter interaction\cite{jo}, which is 
known as the MSW effect\cite{wf5,ms}.

\section{An Explicit Example}

In this section we specialize the main formulae of 
the previous sections to the sequential decay
\begin{equation}
\Lambda_b \to (p\pi^-)_{\Lambda} (\mu^+\mu^-)_{J/\psi},
\end{equation} 
a particular case of decays (\ref{declb}). In this situation,
the differential decay width and the asymmetries can be 
expressed as functions of the components along the unit
vectors (\ref{unvc}) of the polarization vector 
${\vec{\cal P}}$ of the $\Lambda_b$ resonance. Indeed, 
in the canonical frame, the $\Lambda_b$ density 
matrix $\rho^{(0)}$ reads as  
\begin{equation}
\rho^{(0)}_{\pm\pm} = \frac{1}{2}\pm {\cal P}_z,
~~~~~~~~ \rho^{(0)}_{\mp\pm} = 
{\cal P}_x \pm i{\cal P}_y.
\end{equation} 
On the other hand, the unit vectors (\ref{unvc})
can be expressed as functions of the angles 
$\theta$ and $\phi$ in the canonical frame:
\begin{eqnarray}
{\hat{\bf e}}_L &\equiv& (sin\theta cos\phi, sin\theta sin\phi, cos \theta),
\\
{\hat{\bf e}}_T &\equiv& (-sin\phi, cos\phi, 0), ~~~~~~~ \ ~~~~~~~
\\
{\hat{\bf e}}_N &\equiv& (-cos\theta cos\phi, -cos\theta sin\phi, sin\theta).
\end{eqnarray}
The decay is characterized by four amplitudes,
which we denote by 
$A_{1/2, 1}$, $A_{-1/2, -1}$, $A_{1/2, 0}$ and 
$A_{-1/2, 0}$, dropping - from now on - the superscript 
$J$ introduced in eq. (\ref{nmam}). 
The differential decay width reads
\begin{eqnarray}
\Gamma(\Omega) &=& \frac{1}{4\pi}W(1+2{\cal P}_L\Delta G_L),
~~~~~~~ \ ~~~~~~~ \ 
\\
\Delta G_L &=& |\alpha_{1/2,0}|^2-|\alpha_{-1/2,0}|^2-
|\alpha_{1/2,1}|^2+|\alpha_{-1/2,-1}|^2,
\\
{\cal P}_L &=& {\vec{\cal P}}\cdot{\hat{\bf e}}_L.
~~~~~~~ \ ~~~~~~~ \ ~~~~~~~ \
\end{eqnarray} 
Now we give the expressions of some of the 
asymmetries:
\begin{eqnarray}
\Gamma(\Omega) A^L_{\Lambda}(\Omega) &=& \frac{1}
{2\pi}W\Delta a^{\Lambda}(B_L^+ + 2{\cal P}_L B_L^-);
\ ~~~~~~~ \ ~~~~~~~ \
\\
\Gamma(\Omega) A^N_{\Lambda}(\Omega) &=& \frac{1}{\pi}
W \Delta a^{\Lambda} [\Re(\alpha_{1/2,0}
\alpha^*_{-1/2,0}){\cal P}_N + \Im(\alpha_{1/2,0}
\alpha^*_{-1/2,0}){\cal P}_T];
\\
\Gamma(\Omega) A^T_{\Lambda}(\Omega) &=& \frac{1}{\pi}
W \Delta a^{\Lambda} [\Im(\alpha_{1/2,0}
\alpha^*_{-1/2,0}){\cal P}_N - \Re(\alpha_{1/2,0}
\alpha^*_{-1/2,0}){\cal P}_T];
\\
\Gamma(\Omega) A^{LN}_{\Lambda}(\Omega) &=& 
\frac{3\sqrt{2}}{8\pi}W \Delta a^{\Lambda} 
[\Re (C){\cal P}_N - \Im (C){\cal P}_T];
\ ~~~~~~~ \ ~~~~~~~ \
\\
\Gamma(\Omega) A^{LT}_{\Lambda}(\Omega) &=& 
-\frac{3\sqrt{2}}{8\pi}W \Delta a^{\Lambda} 
[\Im (C) {\cal P}_N + \Re (C){\cal P}_T].
\ ~~~~~~~ \ ~~~~~~~ \
\end{eqnarray} 
Here we have set
\begin{eqnarray}
\Delta a^{\Lambda} &=& \frac{1}{2}
(a^{\Lambda}_+ - a^{\Lambda}_-), \ ~~~~~~~ \ ~~~~~~~ \
\ ~~~~~~~ \ 
\\ 
B_L^{\pm} &=& |\alpha_{1/2,1}|^2\pm|\alpha_{1/2,0}|^2
\mp|\alpha_{-1/2,-1}|^2-|\alpha_{-1/2,0}|^2,
\\
{\cal P}_N &=& {\vec{\cal P}}\cdot{\hat{\bf e}}_N,
\ ~~~~~~~ \ {\cal P}_T = {\vec{\cal P}} 
\cdot{\hat{\bf e}}_T, \ ~~~~~~~ \ ~~~~~~~ \  
\\
C &=&  \alpha_{1/2,1}\alpha^*_{1/2,0}-
\alpha_{-1/2,0}\alpha^*_{-1/2,-1}, \ ~~~~~~~ \ 
 ~~~~~~~ \ 
\end{eqnarray} 
$a^{\Lambda}_{\pm}$  being the quantities (\ref{sdch})
referred to the decay of the $\Lambda$-resonance, 
with positive (+) or negative (-) helicity.

The remaining asymmetries defined in the 
preceding sections vanish, because they are
proportional to the quantity
\begin{equation}
\Delta a^V = \frac{1}{2}(a^V_1 - a^V_{-1}),
\end{equation}
where $V$ indicates the vector meson: this
observable is zero, owing to parity conservation 
in strong and electromagnetic decays. 
However, information that we can extract
from the non-vanishing asymmetries is sufficient
to determining the moduli and relative phases
of the reduced amplitudes. Indeed, the moduli
can be inferred from $\Gamma(\Omega)$ and
$A^L_{\Lambda}(\Omega)$; moreover the relative 
phase between $\alpha_{1/2,0}$ and $\alpha^*_{-1/2,0}$
can be extracted from $A^N_{\Lambda}(\Omega)$ or 
from $A^T_{\Lambda}(\Omega)$; lastly, the sine
and cosine of the other two relative phases are
related, respectively, to $\Im (C)$ and to $\Re (C)$,
whence the two phases can be deduced. 

\section{Relations to Previous Works}

In this section we compare our results with those of other authors.
In particular, as we shall see, our choice of considering 
correlations constructed exclusively from momenta results to be 
rather general. Indeed, such
correlations turn out to be spin dependent, yielding results equivalent
to those which involve one or more spins. A strong indication in this
sense comes from the observation that the asymmetries defined in 
sects. 2 and 3 vanish if both the parent resonance and the decay 
products are spinless. 

\subsection{Hadronic $\Lambda_b$ Decays} 

Let us focus on decays of the type (\ref{sdc}), such that $J$ and $s_a$ 
or/and $s_b$ are nonzero. For example, consider decays (\ref{declb}),
studied by several authors\cite{bdl2,bdl1,arg,aj,cgn} and also detected 
experimentally\cite{cdf3,d03,cdf4,d04}.
We define the following products:
\begin{equation}
{\bf s}_{a(b)}\cdot{\bf e}_N, \ ~~~~ \ ~~~ {\bf s}_{a(b)}\cdot{\bf e}_T, \ ~~~~ \ ~~~
{\bf s}_{a(b)}\cdot{\bf e}_L,
\label{prod}
\end{equation} 
whose mean values are the components of the polarization vector of one of 
the two decay products, $a$ or $b$; we denote such components, respectively,  
as $P_N$, $P_T$ and $P_L$. $P_N$ is T-odd, while $P_T$ and $P_L$ are T-even. 
If both $a$ and $b$ are spinning, one can also define polarization 
correlations, like, {\it e. g.}, 
\begin{equation}
P_{TN} = \langle {\bf s}_a\cdot{\bf e}_T ~ {\bf s}_b\cdot{\bf e}_N\rangle.
 \label{plr0} 
\end{equation}
This is the case  of the decays (\ref{declb}), if a vector meson $V$ 
is involved. Here the normal component 
$P_N$ of the polarization of the final fermion reads as\cite{daj1}
\begin{equation}
\Gamma(\Omega)P^{\Lambda}_N(\Omega) \ = \ {\frac{1}{4\pi}}(G^{\Lambda}_N \ + \  
{\Delta}G^{\Lambda}_N  {\cal P}_L), \label{pn0} 
\end{equation}
where 
\begin{eqnarray}
G^{\Lambda}_N \  & = &  \Re{\Big( A_{1/2, 1}{A^{\star}_{-1/2, 0}} + A_{-1/2, 
-1}{A^{\star}_{1/2, 0}} \Big)}, \\
\Delta G^{\Lambda}_N   & = & \ -2\Re{\Big( A_{1/2, 1}{A^{\star}_{-1/2, 0}} - 
A_{-1/2, -1}{A^{\star}_{1/2, 0}} \Big)}. \label{dgn} 
\end{eqnarray}
As regards the polarization correlation $P_{TN}$, 
we have\cite{daj1}
\begin {equation}
\Gamma(\Omega)P_{TN}(\Omega) = {\frac{1}{4\pi\sqrt{2}}}(\Delta G_{TN} \ + \ G_{TN} {\cal P}_L), \label{db2} 
\end{equation}
with
\begin {eqnarray}
G_{TN} & = & 2\Im{(A_{-1/2,-1} A^{\star}_{1/2,0} + A_{1/2,1} 
A^{\star}_{-1/2,0})}, 
\label{gtn}\\
\Delta G_{TN} & = & \Im{(A_{-1/2,-1} A^{\star}_{1/2,0} 
- A_{1/2,1} A^{\star}_{-1/2,0})}. \label{dpol}
\end{eqnarray}
As can be seen, such polarization formulae consist of linear 
combinations of the parameters which appear in the 
single and double asymmetries defined in sects. 2 and 3. Also 
the other components of the polarization vector and the 
polarization correlations are related to our present results 
in a similar way\cite{daj1}: see also sect. 6. It is reasonable 
to think that this result is not peculiar to the decay considered; 
we conjecture that the asymmetries defined in the present paper 
provide, in principle, information on the components of the 
polarization vector and polarization correlations. Even, the 
results of our analysis may be quite suitable for determining 
the components of the polarization vector of a decay product 
which carries a spin greater than 1/2; indeed, it is not so 
easy to measure directly this polarization.

\subsection{$B_{(s)} \to V_1 V_2$ Decays}

As we have seen in sect. 3, double asymmetries include  
decays of the type (\ref{bvv}), for which our analysis
yields the following T-odd observables:
\begin{equation}
\Im\large(\alpha_{11}\alpha^*_{00}\large) 
\ ~~~~ \ {\mathrm and} \ ~~~~ \
\Im\large(\alpha_{-1-1}\alpha^*_{00}\large).
\label{trvo}
\end{equation}
A linear combination of these terms corresponds to
the usual T-odd quantity\cite{dt,dl,kli,dil,ddu}
\begin{equation} 
\Im\large(A_{\perp}A_0^*\large), 
\end{equation}
where the $A$'s are the decay amplitudes in the transversity 
representation. In fact, we have the following relations:
\begin{equation}
A_0 = F\alpha_{00}, ~~~~ A_{\parallel}\ = \frac{1}{\sqrt{2}}
F(\alpha_{11}+\alpha_{-1-1}) 
~~~~ A_{\perp} = \frac{1}{\sqrt{2}}F(\alpha_{11}-\alpha_{-1-1}),
\end{equation}
with
\begin{equation}
F^2 = \sum_{\lambda}|A_{\lambda}|^2 =    |A_0|^2+|A_{\parallel}|^2+|A_{\perp}|^2.
\end{equation}
Moreover, for the decays considered, eq. (\ref{epsi3}) 
yields the following TRV observables:
\begin{equation}
\Im\large(\alpha_{11}\alpha^*_{00}-
{\bar\alpha}_{-1-1} {\bar\alpha}^*_{00}\large) 
\ ~~~~ \ {\mathrm and} \ ~~~~ \
\Im\large(\alpha_{-1-1}\alpha^*_{00}-
\bar{\alpha}_{11} \bar{\alpha}^*_{00}\large),
\label{trv1}
\end{equation}
which are linearly related to $\Im\large(A_{\perp}A_0^*\large)$,
to $\Im\large(A_{\parallel}A_0^*\large)$ and to their 
CP-conjugated quantities.

Some remarks are in order.

a) It is worth stressing that also in this case our method yields 
results which turn out to be very similar to those found by the 
authors just mentioned, although they adopt the T-odd product 
\begin{equation}
{\bf p}_a \cdot {\bf s}_a \times {\bf s}_b.
\end{equation}

b) The usual observables $f_L$ and $f_T$\cite{dt,bek,ch1,dil,bry,ddu}, 
concerning respectively longitudinal and transverse 
polarization of the vector mesons in a decay (\ref{bvv}), 
and used for testing the SM\cite{dil,bry,ka}, are simply 
related to our "reduced" amplitudes: 
\begin{equation}
f_L = |\alpha_{00}|^2, ~~~~~ f_T =  |\alpha_{11}|^2
+ |\alpha_{-1-1}|^2.    
\end{equation}

c) The T-odd observable\cite{dl} 
$\Im\large(A_{\perp}A_{\parallel}^*\large)$ does not 
appear among the parameters of our analysis. This
is due to the fact that we have integrated over
some angular variables in the sequential decay,
whereas one usually adopts a full angular 
analysis\cite{ba1,be2,bpk2,bpk3,be1}, in terms of two
polar angles and an azimuthal one\cite{ddr,cw,ss}.
But if, as discussed in sects. 4 and 6, our method 
allows to extract the moduli and relative phases of 
{\it all} of the decay amplitudes, the above mentioned 
term can be deduced indirectly, as well as those which 
do not appear as parameters of our analysis. Our 
procedure may be suitably followed in cases where 
the statistics is not so rich as in Babar\cite{bpk2} 
and Belle\cite{bpk3} experiments. 

d) The TRV observables proposed by previous 
authors, {\it e. g.},
\begin{equation}
\Im\large(A_{\perp}A_0^*+{\bar A}_{\perp}{\bar A}_0^*\large),
\label{autr} 
\end{equation}
involve just the decay amplitudes. But we observe 
that the proportionality constant ${\bar F}$ relative to
the CP-conjugated decay is different than $F$. This
implies that the observable (\ref{autr}) is substantially
different than (\ref{trv1}). More generally, it 
confirms once more that the TRV observables defined 
in subsects. 5.1 and 5.2 are really different 
than those proposed in the literature. 

\subsection{Detection of Top Decays and New Particle Decays}

Also some of the top decays\cite{at,avl,hmy} and of the possible 
new particle sequential decays\cite{moo,els,lnk,bchk} are studied 
by means of triple products. In particular, it is pointed 
out that correlations from three or more momenta\cite{moo} appear 
to be more appropriate for experimental detection and that the 
maximal asymmetry is obtained by operating in the rest frame 
of the decaying particle\cite{moo}.

\section{Tests of Self-Consistency and of SM}

Here we propose some tests, based on the observables 
illustrated in sections 2, 3 and 5. Moreover,
as shown in sect. 4 and in Appendix, the analysis that 
we have illustrated allows to determine quantities of 
the type
\begin{equation}
|A_1|^2, ~~~~~ |A_2|^2, ~~~~~ \Re (A_1A_2^*), ~~~~~ 
\Im (A_1A_2^*),
\end{equation}
where $A_1$ and $A_2$ are two different helicity 
amplitudes of a given decay mode. Now we propose 
various kinds of tests, based on the knowledge of
such parameters and of others, related to them. 

\subsection{Self-consistency Tests}

A first kind of tests is a self-consistency one.
We suggest to exploit the identities of the type
\begin{equation}
[\Re (A_1A_2^*)]^2+[\Im (A_1A_2^*)]^2 = |A_1|^2|A_2|^2,
\end{equation}
as constraints on the parameters to be extracted 
from data. 

\subsection{Tests of Moduli and Phases}
As told in the introduction,
it is always convenient to look for observables whose
values are small according to the SM, since they may give a
clear indication of NP. In particular, it appears 
suitable to study decays such that the SM predicts negligibly
small TRV or/and direct CPV.

Some tests in this sense were already proposed in a preceding 
paper\cite{ajd}. For instance, we suggested to consider the 
asymmetries
\begin{equation}
{\cal A}_{CP} = 
\frac{\Phi_{\lambda_a \lambda_b}-{\bar\Phi}_{-\lambda_a -\lambda_b}}
{\Phi_{\lambda_a \lambda_b} +{\bar\Phi}_{-\lambda_a -\lambda_b}} 
\ ~~~ {\mathrm and} ~~ \
{\cal A}_M = \frac{|A_{\lambda_a \lambda_b}|^2-|\bar{A}_{-\lambda_a 
-\lambda_b}|^2}{|A_{\lambda_a \lambda_b}|^2+|\bar{A}_{-\lambda_a -\lambda_b}|^2}. 
\label{asy}
\end{equation}
Here $\Phi_{\lambda_a \lambda_b}$ is the relative phase 
of the amplitude $A_{\lambda_a \lambda_b}$  to a fixed  
one, taken as a reference. 

Aside from that,  as already observed by various authors 
(see, {\it e. g.}, ref. \cite{va,dt,bdl2,dl,kli}), a 
hierarchy among the CP-odd or T-odd observables 
has to be established, according to their sensitivity to NP. 
For example, consider the interference term $A_1A_2^*$. In 
order to evaluate qualitatively the behavior of this term, 
we assume a quite simplified model, such that each amplitude 
is of the form 
\begin{equation}
A = Te^{i\omega_T}+Pe^{i(\omega_P+\delta_P)}.     \label{t+p}
\end{equation}
Here  we have omitted, for the sake of simplicity, the 
index 1 or 2; moreover the former term corresponds to the 
tree contribution and the latter to the penguin graph; in 
particular, the $\omega$'s (assumed to be helicity independent)
and the $\delta$'s are, respectively, the weak and the strong 
phases. As usual, we assume $\delta_P$ to be attributed
exclusively the absorptive part of the penguin 
diagram\cite{wf1}. Now we substitute eq. (\ref{t+p}) into 
the interference term and compare it with its CP-conjugated 
quantity. As a result we get 
\begin {eqnarray}
\Re(A_1A_2^*- {\bar A}_1{\bar A}_2^*) & = & -2sin\Delta\omega
(|P_1| |T_2| sin\delta_{P1}+ |P_2| |T_1| sin\delta_{P2}), \label{rph}
\\
\Im(A_1A_2^*- {\bar A}_1{\bar A}_2^*)& = & 2sin\Delta\omega
(|P_1| |T_2| cos\delta_{P1}- |P_2| |T_1|cos\delta_{P2}), \label{iph}
\end{eqnarray} 
where $\Delta\omega = \omega_P-\omega_T$.
Then we conclude that, if the strong phase is small, 
the  real part of the interference term is surely 
negligible, while the imaginary part may be sizeable, 
provided the difference $||P_1| |T_2|-|P_2| |T_1||$ is 
sufficiently large\cite{wf1}. Therefore the imaginary
parts of the complex quantities (\ref{epsi}) and
(\ref{epsi3}), as well as the observable (\ref{lsc}), 
appear favourite in the search for possible clues of NP. 

\subsection{Helicity tests}

Here we suggest an interesting test of the SM, 
as a generalization of the one proposed in 
ref.\cite{ddu}. This test is not connected to interference 
terms, but it is an important byproduct of the helicity 
representation, used in our treatment. Indeed, 
the SM predicts positive helicity
amplitudes to be strongly suppressed with respect to
negative helicity ones, if heavy quarks are
involved in a decay. This is verified at an
empirical level, although theoretical arguments
are not so sound\cite{bry}. Then violation
of this inequality  may be an indication of NP.
Incidentally, if the decay products are spinning,
and at least one of them is a vector meson, it is interesting 
to compare the 0-helicity amplitudes with the 
corresponding ones with $\pm 1$-helicity amplitudes.
Factorization, generally satisfied by the tree contribution,
predicts that the 0-helicity amplitudes are much 
greater\cite{bry}. On the contrary, the penguin contribution,
for which factorization fails, yields comparable
0- and $\pm 1$-helicity amplitudes\cite{bry}. Therefore 
the ratios between such amplitudes indicate the relative
weight of penguin to tree term. It is worth noting that 
such kinds of tests can be suitably performed by means of 
the fake T-odd observables
introduced in sect. 5, especially if the weak phases
are negligibly small\cite{ddu}.

\subsection{$\Lambda_b$ and $\Lambda_c$ Decays}

It is useful to consider the application of the previous 
tests to decays of heavy baryons, for example to 
those described by eq. (\ref{declb}), in view of the 
forthcoming LHCb data. Contributions in this 
sense, especially if one of the decay products is a vector 
meson $V$, have been 
already given\cite{bdl2,bdl1,ajd,arg,aj,cgn,daj1,ajt}. 
According to the helicity 
tests, we expect 
$|A_{1/2, 1}| << |A_{-1/2, -1}|$ and 
$|A_{1/2, 0}| << |A_{-1/2, 0}|$, the $A$'s being
the decay amplitudes introduced in sect. 6.

Moreover the ratios $|A_{-1/2, -1}/A_{-1/2, 0}|$ and 
$|A_{1/2, 1}/A_{1/2, 0}|$ may give indications on the 
proportions with which penguin and tree diagram 
contribute to the decay amplitudes. Let us consider 
the case of $V = J/\psi$\cite{daj1}. Assuming 
amplitudes of the type (\ref{t+p}), the SM predicts
\begin{equation}
T e^{i\omega_T} = V_{bc} V_{sc}^*T', \ ~~~~ \ ~~~~ \  
P e^{i(\omega_P+\delta_P)} = 
\sum_q V_{bq} V_{sq}^*P'_q. \label{lblj}
\end{equation}
Here the $V$'s are elements of the CKM matrix, $T$, $P$  
and $T'$  are real positive numbers and $q = u,c,t$. 
Taking into account the orthogonality condition 
\begin{equation}
\sum_q V_{bq} V_{sq}^* = 0, \label{ort}
\end{equation}
and the fact that $V_{bu}V_{su}^*$ is negligible 
in comparison with the other two terms, we conclude 
that the phase difference $\Delta\omega$ is quite 
small according to the SM. However, recent 
experimental results, concerning both $B_s-{\bar B}_s$ 
mixing\cite{cdf,d0} and direct CP violation asymmetry 
in $B\to K\pi$ decays\cite{bpk1,bpk2,bpk3}, indicate 
that this phase could be considerably larger. 
Therefore it is worth applying our analyses to these 
decay modes. Incidentally, we signal that a commonly 
used model\cite{mrz,hig,aj} predicts $P/T$ $\sim$
0.134\cite{aj}.    

Even more intriguing is the case of $V = \rho^0, 
\omega$\cite{ajt}, or the decay $\Lambda_b\to \Lambda 
\pi^0$. Indeed, here the penguin 
contribution is greater than the tree; therefore
the interference between the two terms, which 
includes the CP-violating phase, is quite relevant. 
Moreover, the diagrams involved are quite 
similar to those which occur in the decays 
$B\to K\pi$\cite{bpk1,bpk2,bpk3}. Therefore it is 
worth doing efforts for revealing such decay modes 
of $\Lambda_b$ and for applying to them 
the tests suggested in this section.

As a further example of decay, consider\cite{kld}
\begin{equation}
\Lambda_c\to \Lambda \pi^+. \label{llp}
\end{equation}
In this case no CP violation has been observed\cite{kld};
however an analysis of the triple product asymmetry is 
suggested, since one has to do with a decay where 
the relative strong phase of the two amplitudes is quite small. 
Obviously, our method appears to be appropriate
also in this case.

\section{Conclusions}

We have elaborated some methods for analyzing hadronic 
sequential two-body decays involving more spinning particles. 
In particular, we have suggested to investigate several 
distributions, based on T-odd and T-even, simple 
or double correlations. Unlike other authors\cite{dqs}, who 
try to disentangle different CP eigenstates, we exploit just 
the interference between such eigenstates. 
The decays considered offer a richer range of observables 
sensitive to CPV and TRV. They may also help to find hints 
to NP, provided we focus preferably on observables for which 
the SM predicts quite small values. Now we exhibit the highlights 
of our analysis.

a) Our main result is that, given a set of data concerning the 
above mentioned decays, one can always infer a set of TRV 
observables, even after integrating over all variables, except 
for the direction of one of the two decay products in the initial 
decay. We stress that this would not be possible if less than two 
of the particles involved in this decay are spinning. Some of 
these observables are especially sensitive to TRV and to 
possible clues beyond the SM. Among them we signal the imaginary 
parts of the interference terms like (\ref{epsi}) and
(\ref{epsi3}), which are rather small according to the SM 
predictions. In particular, the study of the interference terms 
in the decay modes $\Lambda_b \to \Lambda \rho^0,\omega$ 
appears very interesting in this sense.   

b) We have shown that the T-odd correlations based just on momenta 
(three or more), which may be more easily adopted in an experiment, 
provide the same results as those involving one or more spins.

c) Thirdly, T-even observables - quite helpful, although a bit 
neglected in the  current literature - may be combined with their 
CP-conjugated ones, so as to include either possible CPT-odd terms 
or products of a fake T-odd amplitude, times a real TRV one. 
These observables, as well as fake T-odd ones, may be used to set 
constraints on the T-odd terms caused by strong interactions, 
typically spin-orbit ones.

d) Our treatment recovers, as particular cases, some 
methods suggested by other authors. For example, the analysis 
of the helicity amplitudes, rather efficient as a test of the SM, 
is an important byproduct of our choice of the helicity 
representation. Furthermore, new TRV observables have been picked 
out, the "reduced" amplitudes, which generalize the polarization 
parameters $f_L$ and $f_T$, used for vector mesons.

e) Lastly, our method may allow, at least in some cases, to 
determine the moduli and the relative phases of the decay amplitudes.  
These quantities allow, in turn, to infer further, especially 
sensitive TRV observables and the polarization vectors of the decay 
products, otherwise difficult to determine for, say, a vector meson. 

The analysis proposed here could be usefully applied to forthcoming data, 
especially at LHCb, but also in experiments where the available statistics 
is not so rich and abundant.

\vskip 0.25in
\centerline{\bf Acknowledgments}
The authors are very grateful to their friend J. Orloff for very useful and 
illuminating discussions.

\vskip 0.30in
\vspace {10pt}

\vskip 0.30in
\setcounter{equation}{0}
\renewcommand\theequation{A. \arabic{equation}}

\appendix{\large \bf Appendix}
                  
We describe a method for extracting from data the moduli squared 
of the helicity decay amplitudes and some interference terms 
between them.                   
The method is based on the moment expansion of the various
distributions defined in the text, that is, $\Gamma(\Omega)$, 
$A_a^N(\Omega)\Gamma(\Omega)$, etc., which we denote 
generically by $F(\Omega)$. Each such distribution contains the
product ${\cal D}^{J*}_{M\Lambda}(\Omega){\cal D}^J_{M'\Lambda'}(\Omega)$
of Wigner ${\cal D}$-matrices. Therefore our starting point is                   
the well-known relation 
\begin{equation}
{\cal D}^{J*}_{M\Lambda}(\Omega){\cal D}^J_{M'\Lambda'}(\Omega)
= \sum_L C^{J~~L~J}_{M'NM} C^{JLJ}_{\Lambda'\nu\Lambda}
{\cal D}^{L*}_{N\nu}(\Omega).
\end{equation}
Here the $C$'s are the usual Clebsch-Gordan coefficients.
Inserting this into $F(\Omega)$, we get an expansion of the type  
\begin{equation}
F(\Omega)= \sum_{LN\nu} H^J_{LN\nu}{\cal D}^{L*}_{N\nu}(\Omega).
\end{equation}
Here 
\begin{equation}
H^J_{LN\nu} = t_{LN}^{J*}f^J_{L\nu},  ~~~~~
t_{LN}^{J*}=\sum_{MM'}\rho^{(0)}_{MM'} C^{J~~L~J}_{M'NM}
\label{mmt}
\end{equation}
and the $f^J_{L\nu}$'s vary from distribution to
distribution, as we shall specify below. They depend 
only on the decay amplitudes, while the coefficients 
$t_{LN}^{J*}$ depend only on the production reaction 
of the parent resonance.

{\bf A.1 - Extracting Moduli of Decay Amplitudes}

Now we give the expressions of the coefficients 
$f^J_{L\nu}$ for three distributions, which depend
solely on the moduli of the decay amplitudes, that 
is, $\Gamma(\Omega)$, $\Gamma(\Omega) A_{a(b)}^L(\Omega)$ 
and $\Gamma(\Omega) A^{LL}(\Omega)$, whose expressions are 
given, respectively, by eqs. (\ref{didw}), (\ref{ltash})
and (\ref{LLan}). We have
\begin{eqnarray}
f^{J(\Gamma)}_{L\nu} &=& {\cal N}_J W \delta_{\nu0} \sum_{\lambda_a,\lambda_b}
|\alpha^J_{\lambda_a\lambda_b}|^2 C^{JLJ}_{\Lambda 0\Lambda}, ~~~~~~~~~ \ ~~~~~~~
\ ~~~~~~ \label{fgm}
\\
f^{J(A_a^{\ell})}_{L\nu} &=& 8\pi{\cal N}_{Ja}W \delta_{\nu0} \sum_{\mu_a>0}
\Delta a^{s_a}_{\mu_a} \sum_{\Lambda} 
\Delta a^J_{\Lambda\mu_a} C^{JLJ}_{\Lambda 0\Lambda},
\label{fda}
\\
f^{J(A^{\ell\ell})}_{L\nu} &=& 16\pi^2{\cal N}_{Jab}W  \delta_{\nu0}\sum_{\mu_a>0}
\sum_{\mu_b>0}\Delta a^{s_a}_{\mu_a}\Delta a^{s_b}_{\mu_b} \sum_{\Lambda}
\Delta^{(2)}_{\Lambda\mu_a\mu_b} C^{JLJ}_{\Lambda 0\Lambda}.
\label{fdd}
\end{eqnarray}
Here the various symbols introduced are defined in the text. We
just reproduce those expressions which contain the "reduced" 
amplitudes $\alpha^J_{\lambda_a\lambda_b}$:
\begin{eqnarray}
\Delta a^J_{\Lambda\mu_a} &=& \frac{1}{2}\sum_{\lambda_a > 0}
\delta^{s_a}_{\lambda_a\mu_a}(|\alpha^J_{\lambda_a\lambda_b}|^2-
|\alpha^J_{-\lambda_a\lambda'_b}|^2),
\\
\Delta^{(2)}_{\Lambda\mu_a\mu_b} &=& \sum_{\lambda_a}
|\alpha^J_{\lambda_a\lambda_b}|^2\delta^{s_a}_{\lambda_a\mu_a}
\delta^{s_b}_{\lambda_a\mu_b}. ~~~~~~~~~ \ ~~~~~~~ \label{D2}
\end{eqnarray} 
Moreover the upper indices of the coefficients $f^J_{L\nu}$  
refer to the specific distributions considered; in particular, 
the index $\ell$ refers to ''longitudinal asymmetry'', not 
to be confused with the order $L$ of the moment. Lastly, 
the coefficient $f^{J(A_b^{\ell})}_{L\nu}$ can be defined 
analogously to eq. (\ref{fda}), by interchanging index $a$ 
with index $b$. Now we consider the ratios 
\begin{eqnarray}
r_L^{a\ell} &=& H_{LN0}^{J(A_a^{\ell})}/H_{LN0}^{J(\Gamma)},
\label{rla}
\\
r_L^{b\ell} &=& H_{LN0}^{J(A_b^{\ell})}/H_{LN0}^{J(\Gamma)},
\\
r_L^{\ell\ell} &=& H_{LN0}^{J(A^{\ell\ell})}/H_{LN0}^{J(\Gamma)}.
\label{rll}
\end{eqnarray}
Here, again, the upper indices of the $H^J$'s are typical of
the distribution considered. First of all, it is
important to note that the ratios (\ref{rla}) to
(\ref{rll}) are independent of $N$. This constitutes
a check for the moments of the distributions.
Furthermore, by recalling the relations above, 
we obtain the following linear system in the 
moduli squared of the various $\alpha^J_{\lambda_a\lambda_b}$:
\begin{eqnarray}
r_L^{a\ell} \sum_{\lambda_a,\lambda_b} |\alpha^J_{\lambda_a\lambda_b}|^2 C^{JLJ}_{\Lambda 
0\Lambda} &=& 8\pi\frac{{\cal N}_{Ja}}{{\cal N}_J}\sum_{\mu_a>0}\Delta a^{s_a}_{\mu_a} \sum_{\Lambda} 
\Delta a^J_{\Lambda\mu_a} C^{JLJ}_{\Lambda 0\Lambda},  ~~~~~ \ ~~~~~~
\label{eq1}
\\
r_L^{b\ell} \sum_{\lambda_a,\lambda_b} |\alpha^J_{\lambda_a\lambda_b}|^2 C^{JLJ}_{\Lambda 0\Lambda} &=& 8\pi\frac{{\cal N}_{Ja}}
{{\cal N}_J}\sum_{\mu_b>0}\Delta a^{s_b}_{\mu_b} \sum_{\Lambda} 
 \Delta a^J_{\Lambda\mu_b} C^{JLJ}_{\Lambda 0\Lambda}, ~~~~~ \ ~~~~~~
\\
r_L^{\ell\ell} \sum_{\lambda_a,\lambda_b} |\alpha^J_{\lambda_a\lambda_b}|^2 C^{JLJ}_{\Lambda0\Lambda} &=& 16\pi^2\frac{{\cal N}_{Jab}}
{{\cal N}_J}\sum_{\mu_a>0}\sum_{\mu_b>0}
\Delta a^{s_a}_{\mu_a}\Delta a^{s_b}_{\mu_b} \sum_{\Lambda}
\Delta^{(2)}_{\Lambda\mu_a\mu_b} C^{JLJ}_{\Lambda 0\Lambda}. ~~~ \ ~~~
\label{eq3}
\end{eqnarray}
The parameters $\Delta a^{s_{a(b)}}_{\mu_{a(b)}}$, 
relative to the secondary decays, are generally known.
The linear system (\ref{eq1})-(\ref{eq3}) is 
{\it not} homogeneous, owing to the constraint
\begin{equation}
\sum_{\lambda_a,\lambda_b} |\alpha^J_{\lambda_a\lambda_b}|^2=1.
\end{equation}
It is generally over-determined, as we shall see in some examples 
of interest, provided all of the asymmetries are nontrivial.

{\bf A.2 - Inferring Relative Phases of Decay Amplitudes}

The relative phases of the decay amplitudes may
be determined starting from the other distributions
considered in the text, where interference terms between
amplitudes with different helicities are involved. Here 
we limit ourselves to two of the T-odd distributions, 
but the method we suggest can be extended to any 
distribution of sects. 2 and 3. 

According to our formalism, the moments related to the distribution 
$\Gamma(\Omega)A_{ab}^{NT}(\Omega)$ (see the first eq. (\ref{mmt})) 
yield the following decay coefficients $f^J_{L\nu}$:  
\begin{eqnarray}
f^{J(NT)}_{L\nu}&=&  {\cal N}_{Jab}W \sum_{\mu_a,\mu_b} C^{JLJ}_{\Lambda'\nu\Lambda}
a^{s_a}_{\mu_a} a^{s_b}_{\mu_b} ~~~~ \nonumber
\\
&\times&\sum_{\lambda_a}\sum_{\lambda_b} \sum'_{\lambda'_a}\sum'_{\lambda'_b} 
\Theta^{s_a}_{\lambda_a\lambda'_a\mu_a} 
\Theta^{s_b}_{\lambda_b\lambda'_b\mu_b}
\frac{16i(-)^{D_a}}{(\lambda_a-\lambda'_a) (\lambda_b-\lambda'_b)}
\nonumber
\\
&\times& \alpha^J_{\lambda_a\lambda_b}\alpha^{J*}_{\lambda'_a\lambda'_b},
~~~~ 
\\
\Lambda &=& \lambda_a-\lambda_b, ~~~~~ \Lambda' = \lambda'_a-\lambda'_b,
~~~~~ D_a = (\lambda_a-\lambda'_a-1)/2.
  \label{rlbp}
\end{eqnarray}
Here, as in the text, the symbol $ \sum'$ means that $ \lambda_a-\lambda'_a$
or $ \lambda_b-\lambda'_b$ is an odd quantity.
Analogously, the distribution $\Gamma(\Omega)A_{ab}^{LN}(\Omega)$
yields
\begin{eqnarray}
f^{J(LN)}_{L\nu}&=& 2\pi{\cal N}_{Jab}W\sum_{\mu_a,\mu_b} C^{JLJ}_{\Lambda'\nu\Lambda}
a^{s_a}_{\mu_a} a^{s_b}_{\mu_b} \nonumber
\\
&\times&\sum_{\lambda_a}\sum_{\lambda_b}\sum'_{\lambda'_b} 
\delta^{s_a}_{\lambda_a \mu_a}\Theta^{s_b}_{\lambda_b\lambda'_b\mu_b}
\frac{4(-)^{D_b}}{\lambda_b-\lambda'_b} \nonumber
\\
&\times&\alpha^J_{\lambda_a\lambda_b}\alpha^{J*}_{\lambda_a\lambda'_b}.
\end{eqnarray}
Moreover $D_b$ = $(\lambda_b-\lambda'_b-1)/2$ and eqs. (\ref{rlbp}) 
hold. Other equations can be obtained from the other distributions 
considered in sects. 2 and 3.
Once the moduli of the decay amplitudes are known, the ratios 
$H^{J(NT)}_{LN\nu}/H_{LN0}^{J(\Gamma)}$, 
$H^{J(LN)}_{LN\nu}/H_{LN0}^{J(\Gamma)}$ and the other
ratios obtained from the various distributions
depend only on the relative phases of such amplitudes. 
It can be checked that, in the cases of interest (see the
next subsection), a complete set of $n-1$ {\it 
independent} relative phases - $n$ being the number of 
helicity amplitudes - is present in the equations that we
obtain from these ratios. Again, if the asymmetries are all 
nontrivial, this system is over-determined in the cases 
of interest.

{\bf A.3 - Some Examples}

We present some cases of interest of two-body decays in Table 1, 
where we give the number of unknowns and
of available equations. As regards relative phases, we have 
taken into 
account only the conditions deriving from the knowledge of 
the moments defined in 
subsect. A.2. 
\begin{table*}[hbt]
\setlength{\tabcolsep}{1.5pc}
\newlength{\digitwidth} \settowidth{\digitwidth}{\rm 0}
\catcode`?=\active \def?{\kern\digitwidth}
\caption{Moduli and relative phases in different channels}
\label{tab:one}

\begin{tabular*}{\textwidth}{@{}l@{\extracolsep{\fill}}rrrrr}
\hline
                 & \multicolumn{1}{r}{Decay ~~~~} 
                 & \multicolumn{1}{r}{$N_m$} 
                 & \multicolumn{1}{r}{$N^e_m$}
                 & \multicolumn{1}{r}{$N_p$} 
                 & \multicolumn{1}{r}{$N^e_p$}
                    \\ 
\hline
& $1\to 0 ~~~ ~~~ 0$    & $ 1 $  & $ 1 $ & $ 0 $ & $ 0 $ \\         
& $0\to 1 ~~~ ~~~ 1$    & $ 3 $  & $ 4 $ & $ 2 $ & $ 2 $ \\
& $0\to 1/2 ~~~ 1/2$    & $ 2 $  & $ 4 $ & $ 1 $ & $ 2 $ \\
& $1/2\to 1/2 ~~~ 0$    & $ 2 $  & $ 7 $ & $ 1 $ & $ 4 $ \\
& $1/2\to 1/2 ~~~ 1$    & $ 4 $  & $ 7 $ & $ 3 $ & $ 4 $ \\
& $3/2\to 1/2 ~~~ 1$    & $ 6 $  & $13 $ & $ 5 $ & $ 8 $ \\
\end{tabular*}
\end{table*}

Here $N_{m(p)}$ is the number of  moduli (relative phases) of
of the decay amplitudes in the various channels and $N^e_{m(p)}$
the respective numbers of equations.


\end{document}